\documentclass[aps,onecolumn,12pt]{revtex4}
\usepackage{amsmath}
\begin{document}
\title{Conformal Invariance and the Metrication of the Fundamental Forces}
\author{Philip D. Mannheim}
\affiliation{Department of Physics, University of Connecticut, Storrs, CT 06269, USA.
email: philip.mannheim@uconn.edu}
\date{March 28, 2016}
\begin{abstract}
We revisit Weyl's metrication (geometrization) of electromagnetism. We show that by making Weyl's proposed geometric connection be pure imaginary, not only are we able to metricate electromagnetism, an underlying local conformal invariance makes the geometry be strictly Riemannian and prevents observational gravity from being complex. Via torsion we achieve an analogous metrication for axial-vector fields. We generalize our procedure to Yang-Mills theories, and achieve a  metrication of all the fundamental forces. Only in the gravity sector does our approach differ from the standard picture of fundamental forces, with our approach requiring that standard Einstein gravity be replaced by conformal gravity. We show that quantum conformal gravity is a consistent and unitary quantum gravitational theory, one that, unlike string theory, only requires four spacetime dimensions.
\end{abstract}
\maketitle

\vskip3.0truein

\noindent
Essay written for the Gravity Research Foundation 2016 Awards for Essays on Gravitation.

\newpage
\section{Weyl Connection and Conformal Symmetry}

Shortly after Einstein developed general relativity and metricated gravity, a first attempt at a metrication of the fundamental forces was made by Weyl (for a recent discussion see \cite{Yang2014}). With the Riemann tensor $R_{\lambda\mu\nu\kappa}$ being constructed via the Levi-Civita connection  $\Lambda^{\lambda}_{\mu\nu}=(1/2)g^{\lambda\alpha}(\partial_{\mu}g_{\nu\alpha} +\partial_{\nu}g_{\mu\alpha}-\partial_{\alpha}g_{\nu\mu})$, Weyl gave the electromagnetic  potential $ A_{\mu}$ a geometric structure by augmenting  $\Lambda^{\lambda}_{\mu\nu}$ with the Weyl connection $W^{\lambda}_{\mu\nu}=-g^{\lambda\alpha}(g_{\nu\alpha}A_{\mu} +g_{\mu\alpha}A_{\nu}-g_{\nu\mu}A_{\alpha})$. As constructed, the generalized connection $\tilde{\Gamma}^{\lambda}_{\mu\nu}=\Lambda^{\lambda}_{\mu\nu}+W^{\lambda}_{\mu\nu}$ is quite remarkable as it is left invariant under the local conformal transformation 
\begin{eqnarray}
g_{\mu\nu}(x)\rightarrow \exp[2\beta(x)]g_{\mu\nu}(x),~~~A_{\mu}(x)\rightarrow A_{\mu}(x)+\partial_{\mu} \beta(x), 
\label{M1}
\end{eqnarray}
each with the same $\beta(x)$. Consequently, any generalized Riemann tensor $\tilde{R}_{\lambda\mu\nu\kappa}$ built with $\tilde{\Gamma}^{\lambda}_{\mu\nu}$ would be locally conformal invariant too. However, for this $\tilde{\Gamma}^{\lambda}_{\mu\nu}$ the covariant derivative of the metric is the non-zero  $\tilde{\nabla}_{\sigma}g^{\mu\nu}=-2g^{\mu\nu}A_{\sigma}$, with the geometry being a Weyl geometry rather than a Riemannian one. And with parallel transport then being path (and thus history) dependent, the theory is untenable.

Even though the Weyl theory had to therefore be abandoned, the transformation on $A_{\mu}$ was retained. If we couple a Dirac fermion to both $A_{\mu}$ and a set of vierbeins $V^{\mu}_a$ that are needed to implement local Lorentz invariance, the massless Dirac action takes the form  
\begin{eqnarray}
I_{\rm D}=\int d^4x(-g)^{1/2}i\bar{\psi}\gamma^{\mu}(x)(\partial_{\mu}+\Gamma_{\mu} -iA_{\mu})\psi+{\rm H.~c.}, 
\label{M2}
\end{eqnarray}
where $\gamma^{\mu}(x)=V^{\mu}_a(x)\gamma^a$ and $\Gamma_{\mu}(x)=[\gamma^{\nu}(x),\partial_{\mu}\gamma_{\nu}(x)]/8 -[\gamma^{\nu}(x),\gamma_{\sigma}(x)]\Lambda^{\sigma}_{\mu\nu}/8$ is the Levi-Civita based spin connection. The $I_{\rm D}$ action is locally gauge invariant under $\psi(x)\rightarrow \exp[i\alpha(x)]\psi(x)$, $A_{\mu}(x)\rightarrow A_{\mu}(x)+\partial_{\mu} \alpha(x)$, with the factor $i$ appearing in $\partial_{\mu}-iA_{\mu}$ since even though $\partial_{\mu}$ is real, it is $i\partial_{\mu}$ that is Hermitian. Moreover, as constructed, $I_{\rm D}$ is locally conformal invariant too, as it is left invariant under 
\begin{eqnarray}
&&\psi(x)\rightarrow \exp[-3\beta(x)/2]\psi(x),~~~g_{\mu\nu}(x)\rightarrow \exp[2\beta(x)]g_{\mu\nu}(x),
\nonumber\\
&&V^a_{\mu}(x)\rightarrow \exp[\beta(x)]V^a_{\mu}(x),~~~A_{\mu}(x)\rightarrow A_{\mu}(x). 
\label{M3}
\end{eqnarray}
Unlike in the Weyl case, this time $A_{\mu}(x)$ does not transform at all, something that will prove central in the following. As we see, we are essentially  getting local conformal invariance for free, without asking for it a priori. Moreover, we can interpret $\Gamma_{\mu}(x)$ as the gauge field of local conformal invariance in precisely the same way as $A_{\mu}(x)$ acts as the gauge field of local gauge invariance, with both the real and the imaginary parts of the phase of the fermion being gauged. We thus see that if there are no fundamental mass terms at all and all mass is to come from vacuum breaking, local conformal invariance is a quite natural invariance for physics. And recently both Mannheim \cite{Mannheim1989,Mannheim2006,Bender2008a,Bender2008b,Mannheim2011,Mannheim2011a,Mannheim2011b,Mannheim2012,OBrien2012,Mannheim2012a,Mannheim2014,Mannheim2015} and 't Hooft 
\cite{tHooft2010a,tHooft2010b,tHooft2011,tHooft2015a,tHooft2015b} have been advocating that it should play a prominent role in physics, with 't Hooft even noting \cite{tHooft2015b} that a conformal structure for gravity seems to be inevitable.

\section{Metrication of Electromagnetism}

If we replace $\Lambda^{\lambda}_{\mu\nu}$ by $\Lambda^{\lambda}_{\mu\nu}+W^{\lambda}_{\mu\nu}$ in $\Gamma_{\mu}$ (i.e. set $\partial_{\mu}\rightarrow \partial_{\mu}-2A_{\mu}$ in $\Lambda^{\lambda}_{\mu\nu}$), something quite surprising happens: $W^{\lambda}_{\mu\nu}$ drops out of  $I_{\rm D}$ identically and does not couple to the fermion at all \cite{Hayashi1977}, 
\cite{Mannheim2014}. Thus Weyl's metrication never could have described electromagnetism in the first place. To rectify this, with  $\partial_{\mu}-iA_{\mu}$ replacing $\partial_{\mu}$ in the gauge coupling because of Hermiticity,  analogously in the spin connection we replace $W^{\lambda}_{\mu\nu}$ by  \cite{Mannheim2014}
\begin{eqnarray}
V^{\lambda}_{\mu\nu}=-\frac{2i}{3}g^{\lambda\alpha}(g_{\nu\alpha}A_{\mu} +g_{\mu\alpha}A_{\nu}-g_{\nu\mu}A_{\alpha}),
\label{M4}
\end{eqnarray}
(i.e. $\partial_{\mu}\rightarrow \partial_{\mu}-4iA_{\mu}/3$). Then, on inserting $\Lambda^{\lambda}_{\mu\nu}+V^{\lambda}_{\mu\nu}$ into $\Gamma^{\mu}$, we obtain  \cite{Mannheim2014} none other than the $I_{\rm D}$ action given above,
with metrication of electromagnetism thereby being achieved.  The $I_{\rm D}$  action thus has a dual characterization -- it can be generated via local gauge invariance or via a generalized geometric connection. The two viewpoints are equivalent.

While metrication is achieved, there is an immediate concern, an $\Lambda^{\lambda}_{\mu\nu}+V^{\lambda}_{\mu\nu}$ based $\tilde{R}_{\lambda\mu\nu\kappa}$ would be complex, and its associated gravity would not appear to look anything like normal gravity. However, because of conformal invariance, this turns out not to be the case. Specifically, since, as per (\ref{M3}), $A_{\mu}$ does not transform at all under a local conformal transformation, any $\tilde{R}_{\lambda\mu\nu\kappa}$-based action would not be conformal invariant. The only gravitational action that would be permitted by (\ref{M3}) is the conformal gravity action $I_{\rm W}=-\alpha_g\int d^4x(-g)^{1/2}C_{\lambda\mu\nu\kappa}C^{\lambda\mu\nu\kappa}$ where $\alpha_g$ is dimensionless,  and $C_{\lambda\mu\nu\kappa}$ is the Weyl conformal tensor as evaluated with $\Gamma^{\lambda}_{\mu\nu}$ alone. (Under $g_{\mu\nu}(x)\rightarrow \exp[2\beta(x)]g_{\mu\nu}(x)$ all derivatives of $\beta(x)$ identically drop out of $C_{\lambda\mu\nu\kappa}$.) Invariance under (\ref{M3}) thus forces the gravity sector to be strictly Riemannian, and there is no parallel transport problem. The local conformal structure of $I_{\rm D}$ thus prevents  $V^{\lambda}_{\mu\nu}$ from coupling in the gravity sector, to thereby render the theory viable.

\section{Torsion and the Metrication of the Fundamental Forces}

To generalize metrication to incorporate axial symmetry we allow for torsion and introduce the antisymmetric Cartan torsion tensor $Q^{\lambda}_{\phantom{\alpha}\mu\nu}=\Gamma^{\lambda}_{\phantom{\alpha}\mu\nu}-\Gamma^{\lambda}_{\phantom{\alpha}\nu\mu}=-Q^{\lambda}_{\phantom{\alpha}\nu\mu}$, and the associated contorsion $K^{\lambda}_{\phantom{\alpha}\mu\nu}=(1/2)g^{\lambda\alpha}(Q_{\mu\nu\alpha}+Q_{\nu\mu\alpha}-Q_{\alpha\nu\mu})$. With   $S^{\mu}=(1/8)(-g)^{-1/2}\epsilon^{\mu\alpha\beta\gamma}Q_{\alpha\beta\gamma}$, insertion of $\tilde{\Gamma}^{\lambda}_{\mu\nu}=\Lambda^{\lambda}_{\mu\nu}+V^{\lambda}_{\mu\nu}+K^{\lambda}_{\mu\nu}$ into $\Gamma^{\mu}$ is found \cite{Shapiro2002} to change $I_{\rm D}$ to  $I_{\rm D}=\int d^4x(-g)^{1/2}i\bar{\psi}\gamma^{\mu}(x)(\partial_{\mu}+\Gamma_\mu-iA_{\mu} -i\gamma^5S_{\mu})\psi$. This action is invariant  under $\psi(x)\rightarrow \exp[i\gamma_5 \delta(x)]\psi(x)$, $S_{\mu}(x)\rightarrow S_{\mu}(x)+\partial_{\mu}\delta(x)$, with local axial symmetry thus being metricated too. It is our view that rather than being something arcane, torsion manifests itself as an axial gauge boson, one that would then have escaped detection if it acquires a large enough Higgs mechanism mass. Thus if we seek a metrication of fundamental forces through the Weyl and torsion connections, we are led to a quite far reaching conclusion: not only must the fundamental forces be described by local gauge theories, they must be described by spontaneously broken ones.

The extension to the non-Abelian case is also direct. If for instance we put the fermions into the fundamental representation of  $SU(N)\times SU(N)$ with $SU(N)$ generators $T^i$ that obey  $[T^i,T^j]=if^{ijk}T^k$, replace $A_{\mu}$ by $g_VT^iA^{i}_{\mu}$, replace $Q_{\alpha\beta\gamma}$ by $g_AT^iQ^i_{\alpha\beta\gamma}$, and thus replace $S_{\mu}$ by $g_AT^iS^{i}_{\mu}$ in the connections, we obtain a locally $SU(N)\times SU(N)$ invariant Dirac action of the form \cite{Mannheim2014,Mannheim2015}
\begin{eqnarray}
J_{\rm D}&=&\int d^4x(-g)^{1/2}i\bar{\psi}\gamma^{a}V^{\mu}_a(\partial_{\mu}+\Gamma_{\mu}
-ig_VT^iA^i_{\mu} -ig_A\gamma^5T^iS^i_{\mu})\psi. 
\label{M5}
\end{eqnarray}
As regards the bosonic sector of the theory, given all the local symmetries of $J_{\rm D}$, the action must have the generic conformal gravity plus chiral Yang-Mills form 
\begin{eqnarray}
I_{\rm W}+I_{\rm YM}&=&\int d^4x(-g)^{1/2}[ -\alpha_gC_{\lambda\mu\nu\kappa}C^{\lambda\mu\nu\kappa}-\frac{1}{4}G_{\mu\nu}^iG^{\mu\nu}_i-\frac{1}{4}S_{\mu\nu}^iS^{\mu\nu}_i],
\label{M6}
\end{eqnarray}
with the $\Lambda^{\lambda}_{\mu\nu}$, $V^{\lambda}_{\mu\nu}$, and $K^{\lambda}_{\mu\nu}$ contributions breaking up into three distinct and independent sectors. 

However, rather than postulate the bosonic action we can actually generate it dynamically. Specifically, if we introduce the quantum-mechanical path integral $\int D[\psi]D[\bar{\psi}]D[A^i_{\mu}]D[S^i_{\mu}]D[V^a_{\mu}]\exp(iJ_{\rm D})$ and integrate out the fermions (equivalent to a one loop Feynman diagram), up to logarithmically divergent constants we obtain \cite{tHooft2010a},  \cite{Shapiro2002}, \cite{Mannheim2015} an effective action that is precisely of none other than the form given in (\ref{M6}), with the gravity sector of the action expressly having no internal symmetry dependence. Thus save only for the gravity sector, we recognize (\ref{M5}) and (\ref{M6}) as the standard action used in fundamental physics, only now all generated geometrically. Geometry and conformal invariance thus lead us to (\ref{M5}) and (\ref{M6}). We note that we do not induce either the Einstein-Hilbert $I_{\rm  EH}=-(1/16\pi G)\int d^4x (-g)^{1/2}R^{\alpha}_{\phantom{\alpha}\alpha}$ or the cosmological constant $I_{\Lambda}=-\int d^4x (-g)^{1/2}\Lambda$ actions ($G$ and $\Lambda$ both carry dimension). Rather, we expressly induce $I_{\rm W}$, and in fact cannot avoid doing so. The case for conformal gravity has been made in the references, providing a solution to the cosmological constant problem \cite{Mannheim2011,Mannheim2011a,Mannheim2015} and fits \cite{Mannheim2011b,Mannheim2012,OBrien2012} to 138 galactic rotation curves without dark matter or its 276 (two per galaxy) additional free parameters.

\section{The Unitarity Problem}

Since the conformal gravity action is based on fourth-order derivative equations of motion, it has been thought that the theory would not be unitary. Specifically, if one writes the fourth-order propagator $1/k^4$ as  the limit
\begin{eqnarray}
\frac{1}{k^4}=\lim\limits_{M^2\rightarrow 0}\left[\frac{1}{M^2}\left( \frac{1}{k^2}-\frac{1}{k^2+M^2}\right)\right],
\label{M7}
\end{eqnarray}
the presence of the  minus sign would suggest that the closure relation for the propagator modes would be of the form $\sum |n\rangle\langle n|-\sum |m\rangle\langle m|=I$, and thus lead to negative norm ghost states and violations of unitarity. However, on explicitly constructing the quantum-mechanical Hilbert space, Bender and Mannheim \cite{Bender2008a,Bender2008b} found that the states were not normalizable ($\langle m | m \rangle =\infty$), with the proposed closure relation thus being invalid. Also they found that the associated quantum Hamiltonian was not Hermitian, with the standard Dirac norm not being the appropriate norm for the theory. The Hamiltonian was instead found to be $PT$ symmetric, and that, as is characteristic of $PT$ studies, in order to get normalizable states  one has to continue the field operators into the complex plane. When this is done the $PT$ theory norm that results is positive definite and the theory is unitary. With conformal gravity being renormalizable ($\alpha_g$ being dimensionless), and with it being unitary as well, it is offered as a consistent quantum theory of gravity, one that, unlike string theory, only requires four spacetime dimensions.

That conformal gravity must be unitary can be seen from a consideration of the path integral $\int D[\psi]D[\bar{\psi}]D[A^i_{\mu}]D[S^i_{\mu}]D[V^a_{\mu}]\exp(iJ_{\rm D})$. Since fermion path integration generates the conformal gravity action, and since one cannot change the signs of Hilbert space norms in one perturbative loop, either $J_{\rm D}$ and $I_{\rm W}$ both have ghosts or neither does. But $J_{\rm D}$ is the standard ghost-free action used in fundamental physics. Thus conformal gravity must be ghost free too. To understand how this is achieved, we note that there is a hidden assumption in using the path integral. We are assuming that the path integral measure is real, and that path integration over real gravitational fields exists. However, given that to normalize states we had to continue into the complex plane, to make the path integral exist we equally have to continue the path integral measure for the gravitational fields into the complex plane. (In his study of quantum gravity  't Hooft \cite{tHooft2011} also had to use a complex measure.) When this is done the path integral is well-defined, the theory is unitary, and conformal gravity is consistent. To conclude, we note that the use of local conformal invariance not only leads to a metrication of the fundamental forces, it provides them with a consistent quantum dynamics in the presence of gravity.

\end{document}